\documentclass{pasa}%

\usepackage{graphicx}

\title[A sensitive search for water masers associated with star formation regions in the Local Group galaxy NGC\,6822]{A sensitive search for water masers associated with star formation regions in the Local Group galaxy NGC\,6822}

\author[Flanagan et al.]{C. J. P. Flanagan$^1$, S. P. Ellingsen$^1$, A. A. Cole$^1$
\affil{$^1$ School of Natural Sciences, University of Tasmania, Private Bag 37, Hobart, Tasmania 7001, Australia
}%
}%

\jid{PASA}
\doi{10.1017/pas.\the\year.xxx}
\jyear{\the\year}

\usepackage{aas_macros}
\usepackage{hyperref} 
\hypersetup{colorlinks,citecolor=blue,linkcolor=blue,urlcolor=blue}

\begin{document}

\begin{frontmatter}
\maketitle

\begin{abstract}
We report the results of a sensitive search for water maser emission in the Local Group Galaxy NGC\,6822 with the Karl G. Jansky Very Large Array. The observations provide tentative single-epoch detections of four candidates, associated with two infrared-bright star formation regions (Hubble~I/III and Hubble~IV). The candidate maser detections are all offset from the velocity range where strong emission from H\textsc{i} neutral gas is observed toward NGC\,6822, with the closest offset by $\sim$40 kms$^{-1}$.  Our observations include the location of NL1K, a previous tentative water maser detection in NGC\,6822.  We do not detect any emission from this location with a sensitivity limit approximately a factor of 5 better than the original Sardina Radio Telescope observations.
\end{abstract}

\begin{keywords}
masers -- stars: formation --  galaxies: individual (NGC\,6822)
\end{keywords}
\end{frontmatter}

\section{INTRODUCTION }
\label{sec:intro}

Barnard's Galaxy (NGC\,6822) is a dwarf irregular galaxy at a distance of 479$\pm$15 kpc \citep{2012MNRAS.421.2998F}, with a metallicity of around one-fifth solar \citep{swa16, gar16}. NGC\,6822 is one of the nearest star-forming dwarf galaxies, and is the closest dwarf irregular galaxy that is definitely not in a bound orbit with the Milky Way. NGC~6822 hosts seven major star-forming regions, with a total estimated star formation rate of at least 0.04~M$_{\odot}$~yr$^{-1}$ \citep{jon19}. Similarities between dwarf irregulars and high-redshift galaxies, such as their low metallicities and their lack of large-scale internal structure, mean that despite the mass difference the former is potentially a useful proxy for the latter when investigating galactic evolution.

NGC\,6822 has a compact H\textsc{i} velocity distribution. As presented in \citet{2017MNRAS.472.3761N}, the velocity dispersion of the H\textsc{i} gas peaks at ~10 kms$^{-1}$ in the central region of the galaxy, where the barycentric velocity is around $-$57 kms$^{-1}$, and drops to zero at the edges of the galaxy. The central region has sixteen IR-bright regions identified by \textit{Spitzer} observation (as presented by \citet{2006ApJ...652.1170C}). These regions differ slightly from the historical  categorisation of central NGC\,6822 H\textsc{ii} regions, which was based on the work of \citet{1925ApJ....62..409H}.

Intense, distributed infrared (IR) emission is closely associated with star formation, making the IR-brightness of a region a tracer of the star formation rate (SFR) therein (see \citet{2012ARA&A..50..531K}, particularly section 3.4). Compared to the H$_\alpha$ line, IR emission includes less massive stellar populations - while H$_\alpha$ emission almost exclusively comes from stars less than 10 Myr old, broader IR emission can be contributed by stars more than 100 Myr old. This makes IR emission an imprecise tracer of instantaneous SFRs but a reasonable tracer for star formation activity.

Young star-forming regions are well-traced by three species of interstellar masers within the Galaxy: water, methanol and hydroxyl masers (eg. \citet{2007A&A...476..373F}). Beyond the Milky Way, the occurrence of water masers appears to correlate well with the star formation rates of their hosts, while methanol and OH masers are significantly underabundant in the Magellanic Clouds (MCs) \citep{2008MNRAS.385..948G}. However, the relatively low intensity of methanol and OH masers when compared to the 22 GHz water maser transition means the comparison cannot be made over the entire Local Group with any degree of confidence.  With no previously detected masers of any species in NGC\,6822, luminosity considerations mean that any attempt to create a catalogue of NGC\,6822 masers should start with the 22 GHz H$_2$O transition. The abundance of OH and methanol masers in NGC\,6822 can potentially be derived from targeted observations toward detected water masers in that galaxy, which would allow the effects of large-scale environmental conditions on maser populations to be studied.

Generating a catalogue of masers is also critical to determining precise distances and kinematics of the Local Group through Very Long Baseline Interferometry (VLBI) observations. Water masers are not the ideal target for astrometric VLBI measurements, as their temporal instability (e.g. \citet{2013MNRAS.432.1382B}) often stymies precise tracking and they are rarely simply associated with the systemic motion of their host galaxy. However, water masers are often associated with masers of other species (e.g. hydroxyl and methanol in young star-forming regions), and their strength compared to other species present in the same star-forming regions allows astrometry at greater distances. Such astrometry has already been performed on Milky Way populations (e.g. \citet{2012PASJ...64..136H}), Local Group galaxies M33 \citep{2005Sci...307.1440B} and IC 10 \citep{2007A&A...462..101B}, and distant galaxy NGC 5765b \citep{2016ApJ...817..128G}.

Here we present a sensitive search for water maser candidates within NGC\,6822. The primary motivation of this observation was to identify water maser candidates that could be used to determine the proper motion of NGC\,6822 through future VLBI observations. This is a potential independent check on proper motions from the ESA {\it Gaia} mission \citep{mcc21}, with very different systematics. The data was collected at a single epoch, so the temporal variability of the candidates, and so their feasibility for VLBI programs, cannot be assessed from the data.

\section{OBSERVATION AND DATA REDUCTION}
\begin{figure*}[h]
	\includegraphics[width=\linewidth]{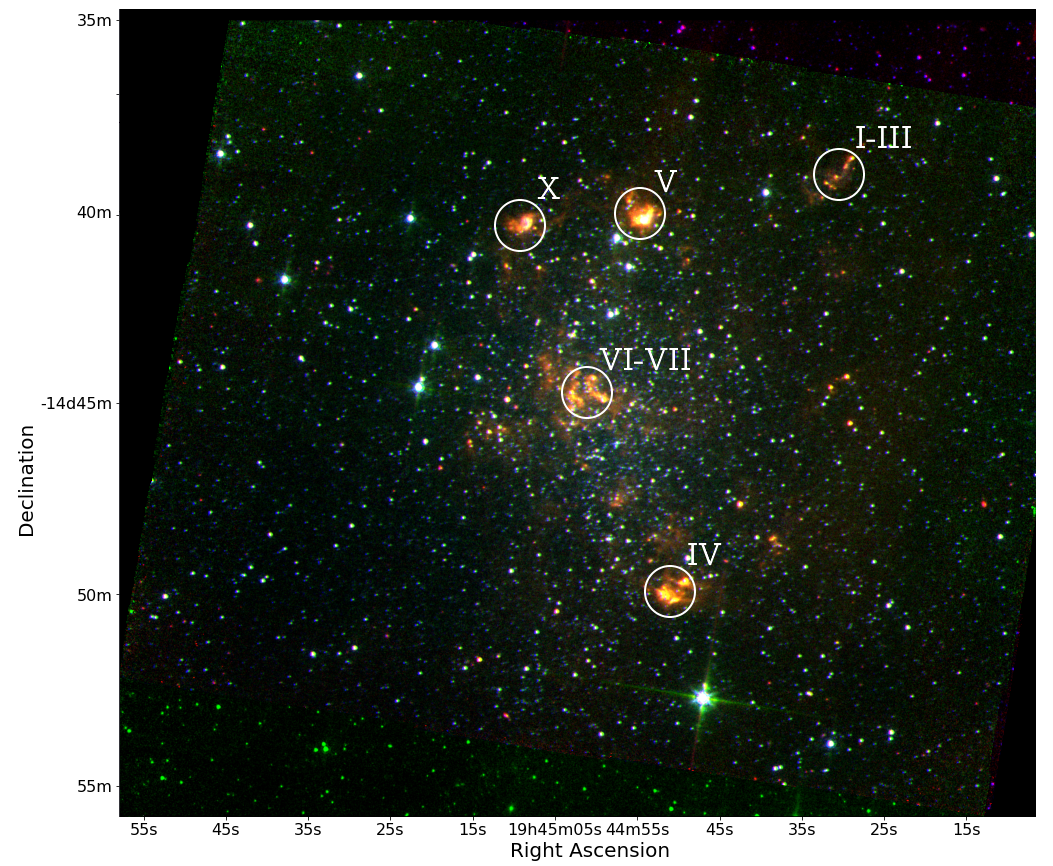}
	\caption{The locations of the five observed fields superimposed on a false-colour image of SIRTF Nearby Galaxies Survey \citep{2003PASP..115..928K}, using their historical labels.}
	\label{fig:3awes}
\end{figure*}

The observation was conducted to search for new detections using four hours of the Karl G. Jansky VLA in D-configuration, between 23.15 and 3.15 UTC on October 12th 2015. Observations targeted the five IR-bright regions of NGC\,6822 shown in Figure \ref{fig:3awes} and used the K-band receiver system that detects left and right circularly polarised signals separately. The observations were split into four blocks, each consisting of sequential observations of each region once followed by pointing offset correction by scanning the nearby quasar J1939-1525.

\begin{table*}[t]
\caption{Regions of NGC\,6822 observed. Cannon labels and emission strengths are from \citet{2006ApJ...652.1170C}; historical labels are the same regions in older literature.}
\centering
\begin{tabular*}{\linewidth}{@{}c\x c\x c\x c\x c\x c\x c\x c\x c\x c\x c\x c@{}}
\hline \hline
 Cannon Label    & Historical Label  &  Right Ascension & Declination & H$_\alpha$ emission & 70 \textit{$\mu$}m emission\\
  & & Hours (J2000) & Degrees (J2000) & 10$^{-13}$ erg s$^{-1}$ cm$^{-2}$ & mJy\\
\hline
3 & Hubble I-III    & 19:44:32.96 & -14:42:10.71 & 110$\pm$20 & 1700$\pm$430\\
7 & Hubble IV     & 19:44:49.47  & -14:52:40.93 & 24$\pm$5 & 2300$\pm$580\\
8 & Hubble V     & 19:44:52.81  & -14:43:04.97 & 120$\pm$25 & 4300$\pm$1100\\
11 & Hubble VI-VII & 19:44:58.59  & -14:47:47.02 & 13$\pm$3 & 2000$\pm$500\\
14 & Hubble X & 19:45:05.01  & -14:43:17.07 & 110$\pm$20 & 2400$\pm$600\\
\hline \hline
\end{tabular*}
\label{tabIRregions}
\end{table*}

The correlator was configured to record 4096 spectral channels over a 64 MHz bandwidth, for both left and right circularly polarised signals. The central frequency for the observations was 22224 MHz, corresponding to a velocity coverage of ~870 kms$^{-1}$ and a spectral resolution of 0.2 kms$^{-1}$. Each individual scan of a region involved 8 minutes of observation, resulting in an average RMS image noise of approximately 2-3 mJy/beam in a 0.5~kms$^{-1}$ channel for the full 32 minutes of data per region.

\begin{figure*}[h]
	\includegraphics{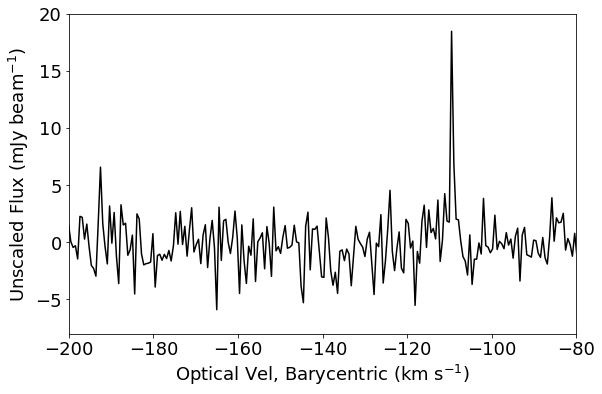}
	\caption{Intensity-velocity spectra of the strongest pixel in the J1944 -1452 1 detection in Field 7, taken on 0.5kms$^{-1}$ channels.}
	\label{fig:J1944_-1452_1SPECTRA}
\end{figure*}

The data were reduced using standard techniques for spectral line reduction in the National Radio Astronomy Observatory (NRAO) Common Astronomy Software Applications (CASA) package. All velocities used in this paper are in the heliocentric reference frame, with an adopted rest frequency of 22.23508 GHz for the 6$_{16}$-5$_{23}$ water maser transition.  For these D-array observations we achieved a synthesised beam of approximately 4 by 4 arcseconds. Data cubes of the entire primary beam (150 by 150 arcseconds) and a restricted velocity range of $\pm$150 kms$^{-1}$ in 0.5 kms$^{-1}$ channels around the systemic velocity of NGC\,6822 (-57 kms$^{-1}$) were created from both the combined scans of each region and separate left- and right-polarised observation data. This velocity range was chosen to cover $\pm$130 kms$^{-1}$ beyond the range observed in the natal gas in NGC\,6822.

\section{RESULTS}

The data cubes were inspected both through manual visualisation within CASA and with the DUCHAMP source-finding package (\citet{2012MNRAS.421.3242W}) to find potential maser emission. For a point to be considered potential emission, it needed to be within 120 arcseconds of the cube center (corresponding with a 2-arcminute primary beam FWHM); have an SNR of at least 6\textit{$\sigma$} in the cube; be present in both left- and right-polarised data cubes with an SNR of at least 4\textit{$\sigma$}; and persist across multiple adjacent channels (to a width of 60kHz). Under these conditions, a total of 69 potential sources were identified.

A properly calibrated image cube where there is no emission is expected to have noise characteristics described by a Gaussian distribution with a mean of zero.  To assess the likely false positive rate of our method for identifying potential maser sources, we repeated our search process on the negative half of the image cube. Both halves contained an overabundance of 5-7\textit{$\sigma$} points compared to a true Gaussian distribution.  A similar number of  'potential emissions' were detected in the negative half as the positive half within this intensity range. However, for 'potential emission' beyond 7\textit{$\sigma$}, the equivalence in the positive and negative halves of the cube disappeared - Field 7 had no negative 'potential emission' points past ~8.5\textit{$\sigma$} while Field 3 had none above ~7\textit{$\sigma$}. While small number statistics means that the four largest excursions from the mean in the image cubes being found entirely in the positive half may be chance, other cubes (of other fields where no water maser candidades were identified) show symmetry between the positive and negative distributions. These new thresholds were taken to be the lower limits above which positive emission could be considered strong maser emission candidates.

Using this approach, from the 69 potential sources we have identified four water maser candidates in NGC\,6822. With a single epoch of observation, their derived positions are accurate to within a few arcseconds. The characteristics of the candidate detections are summarised in Table~\ref{tabcandidate}.

\begin{table*}[t]
\caption{Candidates considered strong or reasonable targets for redetection in any field, arranged by decreasing estimated candidate strength. V$_{bary}$ is the barycentric velocity, while V$_{bary}$ - V$_{sys}$ is the velocity offset from the systemic velocity of NGC\,6822.  Listed corrected intensities have been corrected for primary beam response; line spacing denotes significant gaps in predicted likelihood.}
\centering
\begin{tabular*}{\linewidth}{@{}c\x c\x c\x c\x c\x c\x c\x c\x c\x c\x c\x c@{}}
\hline \hline
 Source  & Candidate & V$_{bary}$   & V$_{bary}$ - V$_{sys}$  & SNR & Right Ascension & Declination & Flux Density & Corrected Intensity \\
  & & kms$^{-1}$ & kms$^{-1}$ & $\sigma$ & Hours (J2000) & Degrees (J2000) & mJy & mJy \\
\hline
J1944 -1452 & 1 & -109.9     & -52.9 & 9.13 & 19:44:52.0 & -14:52:23 & 20.6 & 27.3\\
\\
J1944 -1442 & 2 & -116.9     & -59.9     & 7.86 & 19:44:36.5   & -14:42:13 & 18.2 & 29.9\\
J1944 -1452 & 3 & -169.7      & -112.7      & 9.24 & 19:44:50.6 & -14:52:53 & 20.9 & 21.2\\
\\
J1944 -1442 & 4 & 73.9  & 130.9      & 9.67 & 19:44:36.7   & -14:42:53 & 22.4 & 56\\
\hline \hline
\end{tabular*}
\label{tabcandidate}
\end{table*}

\begin{figure}[b]
	\includegraphics[width=\linewidth]{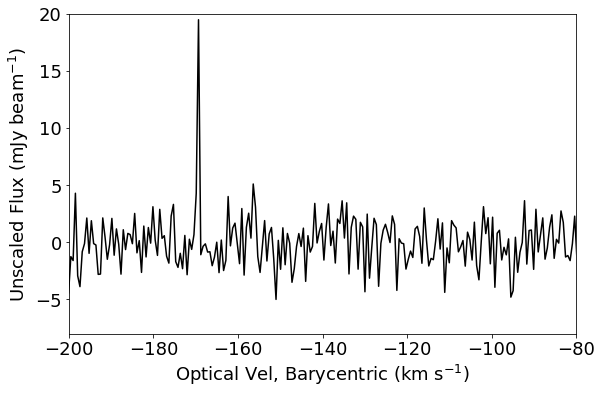}
	\caption{Intensity-velocity spectra of the strongest pixel in the J1944 -1452 3 detection in Field 7, taken on 0.5kms$^{-1}$ channels.}
	\label{fig:J1944_-1452_3SPECTRA}
\end{figure}

\section{DISCUSSION}
The characteristics of the candidate 6$_{16}$-5$_{23}$ water maser detections are listed in Table \ref{tabcandidate}. The four sources have maximum peak flux densities in the range 21-56 mJy and velocity offsets between 40 and 121 kms$^{-1}$ from the gas content dispersion of NGC\,6822. Comparing these to the sample of water masers found by \citet{2014MNRAS.442.2240W} in a follow-up study of the H$_2$O Southern Galactic Plane Survey (HOPS), these offsets lie at extreme but plausible velocities, and correlate with the highest reported intensities of that survey. However, it should be noted that the HOPS survey was not targeted at known high intensity water masers  \citep[e.g.,][]{2010MNRAS.404..779E}, and the candidate detections towards NGC\,6822 are significantly lower in intensity than the brightest Galactic sources would be were they placed at a distance of 479~kpc.

\subsection{Luminosity distribution}

Galactic water maser abundances can be estimated from the empirically determined luminosity function of \citet{1990ApJ...364..513G}

\begin{equation}
\textit{N} = 10^{-0.6(1 + log(L))},
\end{equation}

\noindent
where $N$ is the number of water masers found above some luminosity $L$ in units of $L$$_\odot$. This luminosity function applies for water masers with luminosities between 0.018~$L$$_\odot$ and 1~$L$$_\odot$.  Further work by \citet{2006A&A...457..109B} found good agreement between observed and derived abundances for extragalactic sources beyond this range, provided $N$ is scaled by the relative star formation rate (SFR) of the target galaxy compared to the Milky Way.  The resulting equation being

\begin{equation}
\textit{N} = 10^{-0.6(1 + log(L))} \frac{\text{SFR$_{target}$}}{\text{SFR$_{Galactic}$}}.
\end{equation}

\noindent
Assuming the Milky Way SFR is approximately 1.65 $M_\odot$year$^{-1}$ \citep{2015ApJ...806...96L} and NGC\,6822 has an SFR of approximately 0.04 $M_\odot$year$^{-1}$ \citep{jon19}, $N$ is unity for a luminosity of approximately 0.0002 L$_\odot$.

There are no confirmed detections of masers associated with NGC\,6822; the most recent search being the work of \citet{2020MNRAS.492...45T}, discussed further in section 4.3, which mapped the entire galaxy to a luminosity limit of 0.01 L$_\odot$ (corresponding to a 4$\sigma$ limit of 320 mJy in 1.2kms$^{-1}$ channels) for the 22 GHz water maser line. The lack of any strong candidate detections by \citeauthor{2020MNRAS.492...45T} is consistent with earlier work by \citet{2006A&A...457..109B}, who searched the H$_\alpha$-bright areas of the central region to a luminosity limit of 0.0008 L$_\odot$ (corresponding to a 5$\sigma$ limit of 50 mJy over three 0.7 kms$^{-1}$ channels).  Both previous searches are consistent with Equation (2) in that their sensitivity does not reach the  0.0002 L$_\odot$ threshold below which one or more maser detections are expected. 

Converting between flux density and luminosity for masers depends the channel width of the observations. For example, for the observations reported in this paper we searched for emission in 0.5 kms$^{-1}$ spectral channel, for which a maser with an isotropic luminosity of 1 $L_\odot$ would have a flux density of 379 Jy if it were located at a distance of 479~kpc. For a 0.5 kms$^{-1}$ channel, $N$ = 1 under these conditions for a flux density of 75 mJy; at the intensity of the weakest candidate (21.2 mJy), $N$ is approximately 2. The observations presented in this paper covered a region of NGC\,6822, representing approximately half the galaxy's SFR, so detection of 1-2 masers with a peak intensity of 20 mJy or more is consistent with the Greenhill/Brunthaler luminosity distribution relationship.

However, the extension of the Galactic luminosity function (1) to other galaxies in (2) is not thoroughly empirically vetted; in a search of the Small Magellanic Cloud (SMC) in 2014, \citet{2013MNRAS.432.1382B} report the detection of four water masers above an intensity of 1.3 Jy in 0.5kms$^{-1}$ channels, which corresponds to a threshold of approximately 0.00007 L$_\odot$. At this intensity, $N$ is 2.8, which underpredicts the observed maser abundance by a third.

\subsection{Comparable masers}

Another useful metric for investigating the potential nature of the observed emission is comparison to specific known regions of water maser emission, rather than a statistical comparison to the population of all masers of a particular species.

The Orion KL nebula is the closest high-mass star-forming region, at a distance of approximately 437 pc \citep{2007PASJ...59..897H}. While it has experienced extreme flare events in the past (e.g. \citet{2000AstL...26...34T} and \citet{1989ApJ...338..244G}), achieving flux densities on the order of 10$^6$ Jy, observation between flares shows a relatively quiescent region (e.g. \citet{1998ApJ...493..940G}, where the strongest feature is approximately 14~kJy). Assuming a distance of 479~kpc for NGC\,6822, the weakest of our four candidate detections would have a flux density of approximately 25~kJy were it located at the distance of Orion KL.  So the current observations would not be able to detect emission from a quiescent Orion KL located in NGC\,6822, however, historical flares would be readily detected at levels between 200$\sigma$ and 2500$\sigma$. We should also note that the water masers within Orion KL span a 50~kms$^{-1}$ velocity range centred on the systemic velocity of the quiescent gas, much more compact than that of the candidate detections.

\begin{figure}[b]
	\includegraphics[width=\linewidth]{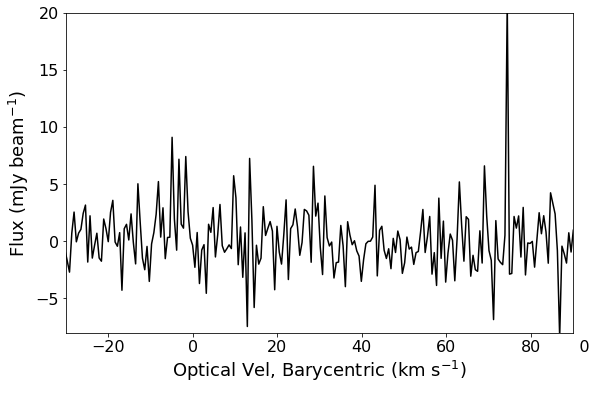}
	\caption{Intensity-velocity spectra of the strongest pixel in the J1944 -1442 4 detection in Field 3, taken on 0.5kms$^{-1}$ channels.}
	\label{fig:J1944_-1442_4SPECTRA}
\end{figure}

A more distant but more active Galactic high-mass star-forming region is the H$_{II}$ region W49, at a parallax distance of approximately 11 kpc \citep{2013ApJ...775...79Z}. The region, W49 North, hosts compact and bright H$_{II}$ sub-regions \citep{2000ApJ...540..308D} with strong water maser activity. In \citet{2004ApJS..155..577M}, a total of 316 water maser components were identified in W49N. 124 of these components (spectral features) could have been detected were they at the distance of NGC\,6822; 23 of them were at least five times as strong as the strongest candidate detection. The aforementioned 124 W49N water maser components spanned a velocity range of 170 to -207 kms$^{-1}$, well beyond that of the natal CO gas, which covers a range of approximately 19 to -3 kms$^{-1}$ (\citet{2009PASJ...61...39M} identify the feature MHH-1 as corresponding to W49N). Even considering only the sources of at least 1 kJy, the velocity range is 56.4 to -113.4 kms$^{-1}$; while more compact, this is still a considerable velocity range, on a similar scale to the candidate detections in NGC\,6822.

Considering sources in galaxies other than the Milky Way, the NGC\,6822 candidate detections are of a similar magnitude to those found by \citet{2013MNRAS.432.1382B} in the SMC, with the strongest candidate at approximately 3.5 Jy (when scaled to the distance of the SMC) and the others at 1.33, 1.71 and 1.87 Jy. The strongest is thus slightly weaker than the strongest NGC\,6822 candidate at 3.4 Jy. 

\begin{figure}[b]
	\includegraphics[width=\linewidth]{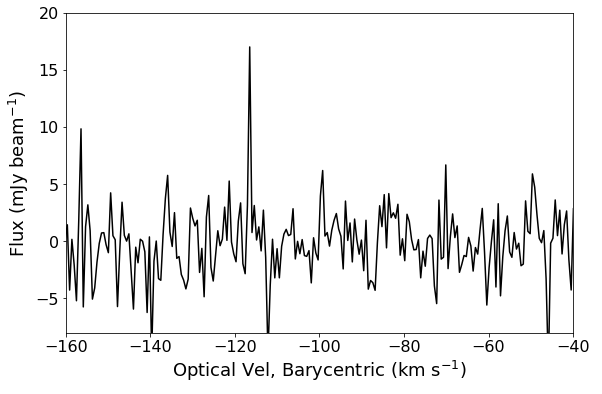}
	\caption{Intensity-velocity spectra of the strongest pixel in the J1944 -1442 2 detection in Field 3, taken on 0.5kms$^{-1}$ channels.}
	\label{fig:J1944_-1442_2SPECTRA}
\end{figure}

All of our NGC\,6822 candidate detections have a very narrow velocity range of a few kms$^{-1}$, a smaller range than typical Galactic masers but similar to those found in the SMC. This is not necessarily indicative of a physical difference, as sensitivity limits mean the full extent of the emission is impossible to determine, but may point to a qualitiative difference in environmental conditions between Galactic masers and those in more gas-rich dwarf galaxies.

The NGC\,6822 candidate detections are thus closest in spread to an atypically active region such as W49 North, but in intensity more similar to the SMC or the brightest edge of more typical Galactic regions such as Orion KL.

\subsection{Previous Tentative detections in NGC6822}

\citet{2020MNRAS.492...45T} report tentative detections of five water masers in NGC 6822, given the placeholder names NL1K through NL5K. These tentative detections have a similar range of line-of-sight velocities to those reported in this paper - offset from the systematic velocity of NGC\,6822 by between 50 km/s and 125 km/s.

Of these five, they identify NL1K and NL5K as being of particular interest, as they are located close to regions of strong continuum emission in NGC\,6822. These correspond to regions 3 and 12 respectively in \citet{2006ApJ...652.1170C} - NL1K thus falls within a region observed by this project. Unfortunately, it does not correspond to any of the candidate water maser detections presented in this paper - the velocity of NLK1 is offset ~80 km/s from the closest candidate in that field (J1944$-$1442). Tarchi et al. also made additional observations with the Sardinia Radio Telescope towards both NL1K and NL5K, but in neither case were they able to confirm the tentative initial detection.

The 7-$\sigma$ detection limit for the current JVLA observations corresponds to a peak flux density of $\sim$20 mJy.  In contrast, the peak flux density of NL1K measured by Tarchi et al. is 543 mJy around late 2015 and 2016, and constrained to have diminished to less than 200 mJy when they attempted to confirm the source in 2019. While significant variability is often found in water masers associated with star formation, the JVLA observations we report here were taken relatively close in time to the initial observations of NL1K.  A source with a peak intensity in excess of 100 mJy would have been readily detected by our observations, strongly suggesting that NL1K was a false positive in the Tarchi et al. search.  This is consistent with Tarchi et al's own estimation of 0.65 false positives at 5 sigma and 72 positives at 4 sigma, as none of the detected sources in NGC\,6822 are above 4.6 sigma - and so the number of tentative detections they report is not higher than would be expected from Gaussian noise.

\section{CONCLUSION}

From a single epoch of targeted water maser observations conducted with the VLA we have identified four water maser candidatess in NGC\,6822.  All of these are below the detection limit of previous surveys. These candidates are high-$\sigma$ detections, but their significant velocity offsets from the gas content of NGC\,6822 and detection at only a single epoch makes it likely that some (perhaps all) are false positives. 

The candidate masers are nearly coincident with two of the bright HII regions in NGC~6822. Candidates 2 and 4 fall just off the western edge of Hubble I/III, northeast of the main bar of the galaxy. Candidates 1 and 3 lie within Hubble IV, in the southern part of the bar. The latter two candidates are directly coincident with areas of high HI column density and radial velocity dispersion. Candidate 3 in particular coincides with a CO detection in field 4 of \citet{2017ApJ...835..278S} and is $\approx$5$^{\prime\prime}$ from a candidate massive Stage~I YSO identified by \citet{jon19} using Spitzer photometry. Water maser emission is {\it not} detected from the two brightest HII regions, Hubble~V and X.

While the radial velocity range overlaps that of Milky Way gas at $l$ = 25$^{\circ}$, it should be noted that the main spiral arms in this direction have $v_{LSR} > 0$, while three of the four candidates in Table~2 have significantly negative velocities; furthermore, at $b$ = $-18^{\circ}$, the sightline toward NGC\,6822 is physically far below the Galactic plane at the typical distances to known spiral arms, making it unlikely that the candidate NGC\,6822 masers are foreground Galactic sources.

If confirmed, they would be consistent with the brightest water masers expected to be present in NGC\,6822 from the luminosity function and will make excellent targets for future VLBI observations to measure the proper motion of NGC\,6822 independently of Gaia \citep{mcc21}.

\begin{acknowledgements}
The Karl G. Jansky VLA is part of the National Radio Astronomy Observatory.

This research has made use of: NASA's Astrophysics Data System Abstract Service; the NASA/IPAC Infrared Science Archive (operated by the Jet Propulsion Laboratory, California Institute of Technology, under contract with the National Aeronautics and Space Administration; the SIMBAD data base, operated at CDS, Strasbourg, France; and data products from the SINGS survey, which is a legacy science program of the Spitzer Space Telescope, funded by the National Aeronautics and Space Administration
\end{acknowledgements}

\begin{figure}[ht]
	\includegraphics[width=\linewidth]{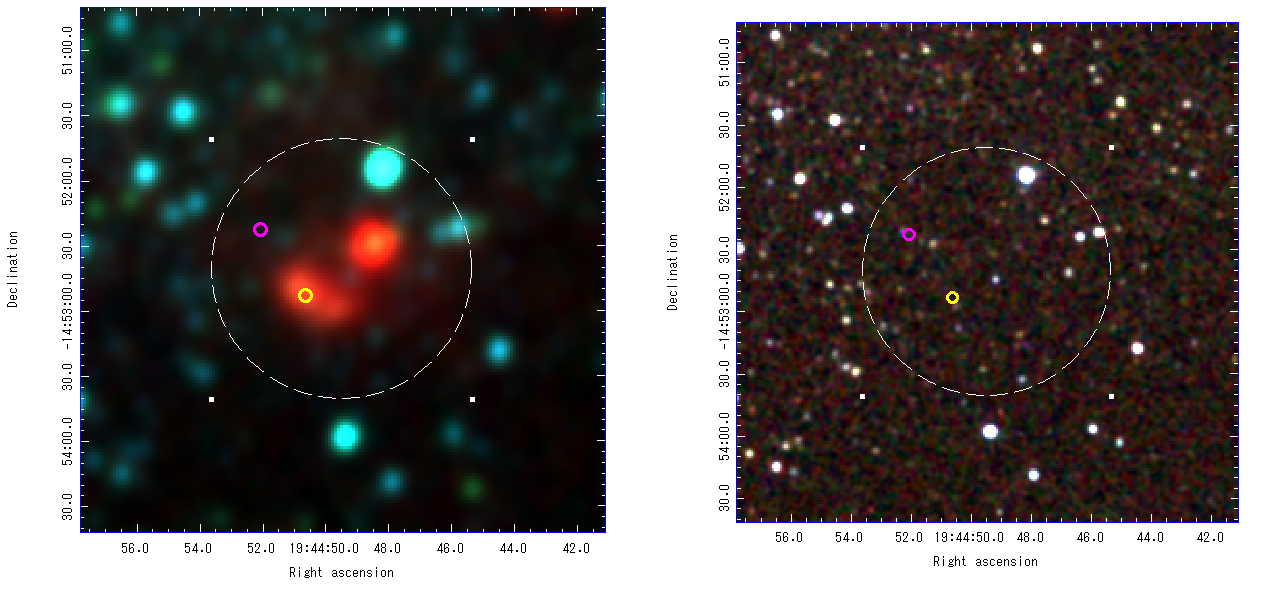}
	\caption{The locations of J1944 -1452 3 and J1944 -1452 1 superimposed on false-colour images of Wide-field Infrared Survey Explorer \citep{2010AJ....140.1868W} 3.4-4.6-12 $\mu$m data (left) and Two-Micron All Sky Survey \citep{2006AJ....131.1163S} J-H-K data (right) . The circles used to indicate the sources J1944 -1452 3 (yellow) and J1944 -1452 1 (purple) have a diameter of 6 arcseconds, approximately the longest dimension of the point-spread function, while the dashed circle corresponds to the primary beam (120 arcsecond diameter).}
	\label{fig:F7WISE}
\end{figure}

\begin{figure}[ht]
	\includegraphics[width=\linewidth]{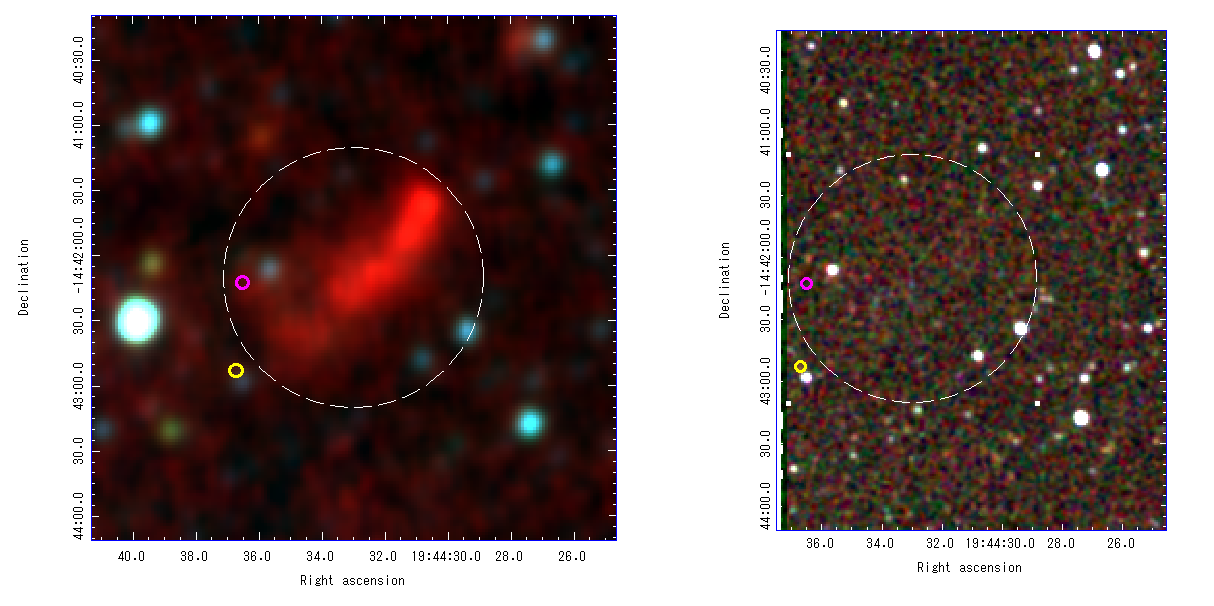}
	\caption{The locations of J1944 -1442 2 and J1944 -1442 4 superimposed on false-colour images of Wide-field Infrared Survey Explorer \citep{2010AJ....140.1868W} 3.4-4.6-12 $\mu$m data (left) and Two-Micron All Sky Survey \citep{2006AJ....131.1163S} J-H-K data (right) . The circles used to indicate the sources J1944 -1442 2 (purple) and J1944 -1442 4 (yellow) have a diameter of 6 arcseconds, approximately the longest dimension of the point-spread function, while the dashed circle corresponds to the primary beam (120 arcsecond diameter).}
	\label{fig:F3WISE}
\end{figure}

\begin{appendix}

\end{appendix}

\bibliographystyle{pasa-mnras}
\bibliography{ngc6822_paper}

\begin{thebibliography}{}
\makeatletter
\relax
\def\mn@urlcharsother{\let\do\@makeother \do\$\do\&\do\#\do\^\do\_\do\%\do\~}
\definecolor{darkblue}{rgb}{0,0,0.597656}
\def\mndoi{\begingroup\mn@urlcharsother \@ifnextchar [ {\mndoi@} {\mndoi@[]}}
\def\mndoi@[#1]#2{\def\@tempa{#1}\ifx\@tempa\@empty \href
  {http://dx.doi.org/#2} {\textcolor{darkblue}{doi:#2}}\else \href
  {http://dx.doi.org/#2} {\textcolor{darkblue}{#1}}\fi \endgroup}
\def\mn@eprint#1#2{\mn@eprint@#1:#2::\@nil}
\def\mn@eprint@arXiv#1{\href {http://arxiv.org/abs/#1} {{\tt arXiv:#1}}}
\def\mn@eprint@dblp#1{\href {http://dblp.uni-trier.de/rec/bibtex/#1.xml}
  {dblp:#1}}
\def\mn@eprint@#1:#2:#3:#4\@nil{\def\@tempa {#1}\def\@tempb {#2}\def\@tempc
  {#3}\ifx \@tempc \@empty \let \@tempc \@tempb \let \@tempb \@tempa \fi \ifx
  \@tempb \@empty \def\@tempb {arXiv}\fi \@ifundefined
  {mn@eprint@\@tempb}{\@tempb:\@tempc}{\expandafter \expandafter \csname
  mn@eprint@\@tempb\endcsname \expandafter{\@tempc}}}

\bibitem[\protect\citeauthoryear{{Breen}, {Lovell}, {Ellingsen}, {Horiuchi},
  {Beasley}  \& {Marvel}}{{Breen} et~al.}{2013}]{2013MNRAS.432.1382B}
{Breen} S.~L.,  {Lovell} J.~E.~J.,  {Ellingsen} S.~P.,  {Horiuchi} S.,
  {Beasley} A.~J.,   {Marvel} K.,  2013, \mndoi [\mnras]
  {10.1093/mnras/stt558}, \href
  {http://adsabs.harvard.edu/abs/2013MNRAS.432.1382B} {432, 1382}

\bibitem[\protect\citeauthoryear{{Brunthaler}, {Reid}, {Falcke}, {Greenhill}
  \& {Henkel}}{{Brunthaler} et~al.}{2005}]{2005Sci...307.1440B}
{Brunthaler} A.,  {Reid} M.~J.,  {Falcke} H.,  {Greenhill} L.~J.,   {Henkel}
  C.,  2005, \mndoi [Science] {10.1126/science.1108342}, \href
  {http://adsabs.harvard.edu/abs/2005Sci...307.1440B} {307, 1440}

\bibitem[\protect\citeauthoryear{{Brunthaler}, {Henkel}, {de Blok}, {Reid},
  {Greenhill}  \& {Falcke}}{{Brunthaler} et~al.}{2006}]{2006A&A...457..109B}
{Brunthaler} A.,  {Henkel} C.,  {de Blok} W.~J.~G.,  {Reid} M.~J.,  {Greenhill}
  L.~J.,   {Falcke} H.,  2006, \mndoi [\aap] {10.1051/0004-6361:20065650},
  \href {http://adsabs.harvard.edu/abs/2006A%26A...457..109B} {457, 109}

\bibitem[\protect\citeauthoryear{{Brunthaler}, {Reid}, {Falcke}, {Henkel}  \&
  {Menten}}{{Brunthaler} et~al.}{2007}]{2007A&A...462..101B}
{Brunthaler} A.,  {Reid} M.~J.,  {Falcke} H.,  {Henkel} C.,   {Menten} K.~M.,
  2007, \mndoi [\aap] {10.1051/0004-6361:20066430}, \href
  {http://adsabs.harvard.edu/abs/2007A%26A...462..101B} {462, 101}

\bibitem[\protect\citeauthoryear{{Cannon} et~al.,}{{Cannon}
  et~al.}{2006}]{2006ApJ...652.1170C}
{Cannon} J.~M.,  et~al., 2006, \mndoi [\apj] {10.1086/508341}, \href
  {http://adsabs.harvard.edu/abs/2006ApJ...652.1170C} {652, 1170}

\bibitem[\protect\citeauthoryear{{De Pree}, {Wilner}, {Goss}, {Welch}  \&
  {McGrath}}{{De Pree} et~al.}{2000}]{2000ApJ...540..308D}
{De Pree} C.~G.,  {Wilner} D.~J.,  {Goss} W.~M.,  {Welch} W.~J.,   {McGrath}
  E.,  2000, \mndoi [\apj] {10.1086/309315}, \href
  {https://ui.adsabs.harvard.edu/abs/2000ApJ...540..308D} {540, 308}

\bibitem[\protect\citeauthoryear{{Ellingsen}, {Breen}, {Caswell}, {Quinn}  \&
  {Fuller}}{{Ellingsen} et~al.}{2010}]{2010MNRAS.404..779E}
{Ellingsen} S.~P.,  {Breen} S.~L.,  {Caswell} J.~L.,  {Quinn} L.~J.,   {Fuller}
  G.~A.,  2010, \mndoi [\mnras] {10.1111/j.1365-2966.2010.16349.x}, \href
  {http://adsabs.harvard.edu/abs/2010MNRAS.404..779E} {404, 779}

\bibitem[\protect\citeauthoryear{{Feast}, {Whitelock}, {Menzies}  \&
  {Matsunaga}}{{Feast} et~al.}{2012}]{2012MNRAS.421.2998F}
{Feast} M.~W.,  {Whitelock} P.~A.,  {Menzies} J.~W.,   {Matsunaga} N.,  2012,
  \mndoi [\mnras] {10.1111/j.1365-2966.2012.20525.x}, \href
  {http://adsabs.harvard.edu/abs/2012MNRAS.421.2998F} {421, 2998}

\bibitem[\protect\citeauthoryear{{Felli} et~al.,}{{Felli}
  et~al.}{2007}]{2007A&A...476..373F}
{Felli} M.,  et~al., 2007, \mndoi [\aap] {10.1051/0004-6361:20077804}, \href
  {https://ui.adsabs.harvard.edu/abs/2007A&A...476..373F} {476, 373}

\bibitem[\protect\citeauthoryear{{Gao} et~al.,}{{Gao}
  et~al.}{2016}]{2016ApJ...817..128G}
{Gao} F.,  et~al., 2016, \mndoi [\apj] {10.3847/0004-637X/817/2/128}, \href
  {https://ui.adsabs.harvard.edu/abs/2016ApJ...817..128G} {817, 128}

\bibitem[\protect\citeauthoryear{{Garay}, {Moran}  \& {Haschick}}{{Garay}
  et~al.}{1989}]{1989ApJ...338..244G}
{Garay} G.,  {Moran} J.~M.,   {Haschick} A.~D.,  1989, \mndoi [\apj]
  {10.1086/167195}, \href {http://adsabs.harvard.edu/abs/1989ApJ...338..244G}
  {338, 244}

\bibitem[\protect\citeauthoryear{{Garc{\'\i}a-Rojas}, {Pe{\~n}a},
  {Flores-Dur{\'a}n}  \& {Hern{\'a}ndez-Mart{\'\i}nez}}{{Garc{\'\i}a-Rojas}
  et~al.}{2016}]{gar16}
{Garc{\'\i}a-Rojas} J.,  {Pe{\~n}a} M.,  {Flores-Dur{\'a}n} S.,
  {Hern{\'a}ndez-Mart{\'\i}nez} L.,  2016, \mndoi [\aap]
  {10.1051/0004-6361/201527265}, \href
  {https://ui.adsabs.harvard.edu/abs/2016A&A...586A..59G} {586, A59}

\bibitem[\protect\citeauthoryear{{Gaume}, {Wilson}, {Vrba}, {Johnston}  \&
  {Schmid-Burgk}}{{Gaume} et~al.}{1998}]{1998ApJ...493..940G}
{Gaume} R.~A.,  {Wilson} T.~L.,  {Vrba} F.~J.,  {Johnston} K.~J.,
  {Schmid-Burgk} J.,  1998, \mndoi [\apj] {10.1086/305161}, \href
  {https://ui.adsabs.harvard.edu/abs/1998ApJ...493..940G} {493, 940}

\bibitem[\protect\citeauthoryear{{Green} et~al.,}{{Green}
  et~al.}{2008}]{2008MNRAS.385..948G}
{Green} J.~A.,  et~al., 2008, \mndoi [\mnras]
  {10.1111/j.1365-2966.2008.12888.x}, \href
  {https://ui.adsabs.harvard.edu/abs/2008MNRAS.385..948G} {385, 948}

\bibitem[\protect\citeauthoryear{{Greenhill}, {Moran}, {Reid}, {Gwinn},
  {Menten}, {Eckart}  \& {Hirabayashi}}{{Greenhill}
  et~al.}{1990}]{1990ApJ...364..513G}
{Greenhill} L.~J.,  {Moran} J.~M.,  {Reid} M.~J.,  {Gwinn} C.~R.,  {Menten}
  K.~M.,  {Eckart} A.,   {Hirabayashi} H.,  1990, \mndoi [\apj]
  {10.1086/169433}, \href {http://adsabs.harvard.edu/abs/1990ApJ...364..513G}
  {364, 513}

\bibitem[\protect\citeauthoryear{{Hirota} et~al.,}{{Hirota}
  et~al.}{2007}]{2007PASJ...59..897H}
{Hirota} T.,  et~al., 2007, \mndoi [\pasj] {10.1093/pasj/59.5.897}, \href
  {https://ui.adsabs.harvard.edu/abs/2007PASJ...59..897H} {59, 897}

\bibitem[\protect\citeauthoryear{{Honma} et~al.,}{{Honma}
  et~al.}{2012}]{2012PASJ...64..136H}
{Honma} M.,  et~al., 2012, \mndoi [\pasj] {10.1093/pasj/64.6.136}, \href
  {https://ui.adsabs.harvard.edu/abs/2012PASJ...64..136H} {64, 136}

\bibitem[\protect\citeauthoryear{{Hubble}}{{Hubble}}{1925}]{1925ApJ....62..409H}
{Hubble} E.~P.,  1925, \mndoi [\apj] {10.1086/142943}, \href
  {https://ui.adsabs.harvard.edu/abs/1925ApJ....62..409H} {62, 409}

\bibitem[\protect\citeauthoryear{{Jones}, {Sharp}, {Reiter}, {Hirschauer},
  {Meixner}  \& {Srinivasan}}{{Jones} et~al.}{2019}]{jon19}
{Jones} O.~C.,  {Sharp} M.~J.,  {Reiter} M.,  {Hirschauer} A.~S.,  {Meixner}
  M.,   {Srinivasan} S.,  2019, \mndoi [\mnras] {10.1093/mnras/stz2560}, \href
  {https://ui.adsabs.harvard.edu/abs/2019MNRAS.490..832J} {490, 832}

\bibitem[\protect\citeauthoryear{{Kennicutt} \& {Evans}}{{Kennicutt} \&
  {Evans}}{2012}]{2012ARA&A..50..531K}
{Kennicutt} R.~C.,  {Evans} N.~J.,  2012, \mndoi [\araa]
  {10.1146/annurev-astro-081811-125610}, \href
  {https://ui.adsabs.harvard.edu/abs/2012ARA&A..50..531K} {50, 531}

\bibitem[\protect\citeauthoryear{{Kennicutt} Robert~C. et~al.,}{{Kennicutt}
  et~al.}{2003}]{2003PASP..115..928K}
{Kennicutt} Robert~C. J.,  et~al., 2003, \mndoi [\pasp] {10.1086/376941}, \href
  {https://ui.adsabs.harvard.edu/abs/2003PASP..115..928K} {115, 928}

\bibitem[\protect\citeauthoryear{{Licquia} \& {Newman}}{{Licquia} \&
  {Newman}}{2015}]{2015ApJ...806...96L}
{Licquia} T.~C.,  {Newman} J.~A.,  2015, \mndoi [\apj]
  {10.1088/0004-637X/806/1/96}, \href
  {https://ui.adsabs.harvard.edu/abs/2015ApJ...806...96L} {806, 96}

\bibitem[\protect\citeauthoryear{{McConnachie}, {Higgs}, {Thomas}, {Venn},
  {C{\^o}t{\'e}}, {Battaglia}  \& {Lewis}}{{McConnachie} et~al.}{2021}]{mcc21}
{McConnachie} A.~W.,  {Higgs} C.~R.,  {Thomas} G.~F.,  {Venn} K.~A.,
  {C{\^o}t{\'e}} P.,  {Battaglia} G.,   {Lewis} G.~F.,  2021, \mndoi [\mnras]
  {10.1093/mnras/staa3740}, \href
  {https://ui.adsabs.harvard.edu/abs/2021MNRAS.501.2363M} {501, 2363}

\bibitem[\protect\citeauthoryear{{McGrath}, {Goss}  \& {De Pree}}{{McGrath}
  et~al.}{2004}]{2004ApJS..155..577M}
{McGrath} E.~J.,  {Goss} W.~M.,   {De Pree} C.~G.,  2004, \mndoi [\apjs]
  {10.1086/424486}, \href
  {https://ui.adsabs.harvard.edu/abs/2004ApJS..155..577M} {155, 577}

\bibitem[\protect\citeauthoryear{{Miyawaki}, {Hayashi}  \&
  {Hasegawa}}{{Miyawaki} et~al.}{2009}]{2009PASJ...61...39M}
{Miyawaki} R.,  {Hayashi} M.,   {Hasegawa} T.,  2009, \mndoi [\pasj]
  {10.1093/pasj/61.1.39}, \href
  {https://ui.adsabs.harvard.edu/abs/2009PASJ...61...39M} {61, 39}

\bibitem[\protect\citeauthoryear{{Namumba}, {Carignan}, {Passmoor}  \& {de
  Blok}}{{Namumba} et~al.}{2017}]{2017MNRAS.472.3761N}
{Namumba} B.,  {Carignan} C.,  {Passmoor} S.,   {de Blok} W.~J.~G.,  2017,
  \mndoi [\mnras] {10.1093/mnras/stx2256}, \href
  {http://adsabs.harvard.edu/abs/2017MNRAS.472.3761N} {472, 3761}

\bibitem[\protect\citeauthoryear{{Schruba} et~al.,}{{Schruba}
  et~al.}{2017}]{2017ApJ...835..278S}
{Schruba} A.,  et~al., 2017, \mndoi [\apj] {10.3847/1538-4357/835/2/278}, \href
  {http://adsabs.harvard.edu/abs/2017ApJ...835..278S} {835, 278}

\bibitem[\protect\citeauthoryear{{Skrutskie} et~al.,}{{Skrutskie}
  et~al.}{2006}]{2006AJ....131.1163S}
{Skrutskie} M.~F.,  et~al., 2006, \mndoi [\aj] {10.1086/498708}, \href
  {http://adsabs.harvard.edu/abs/2006AJ....131.1163S} {131, 1163}

\bibitem[\protect\citeauthoryear{{Swan}, {Cole}, {Tolstoy}  \& {Irwin}}{{Swan}
  et~al.}{2016}]{swa16}
{Swan} J.,  {Cole} A.~A.,  {Tolstoy} E.,   {Irwin} M.~J.,  2016, \mndoi
  [\mnras] {10.1093/mnras/stv2774}, \href
  {https://ui.adsabs.harvard.edu/abs/2016MNRAS.456.4315S} {456, 4315}

\bibitem[\protect\citeauthoryear{{Tarchi} et~al.,}{{Tarchi}
  et~al.}{2020}]{2020MNRAS.492...45T}
{Tarchi} A.,  et~al., 2020, \mndoi [\mnras] {10.1093/mnras/stz3445}, \href
  {https://ui.adsabs.harvard.edu/abs/2020MNRAS.492...45T} {492, 45}

\bibitem[\protect\citeauthoryear{{Tolmachev}}{{Tolmachev}}{2000}]{2000AstL...26...34T}
{Tolmachev} A.~M.,  2000, \mndoi [Astronomy Letters] {10.1134/1.20366}, \href
  {https://ui.adsabs.harvard.edu/abs/2000AstL...26...34T} {26, 34}

\bibitem[\protect\citeauthoryear{{Walsh}, {Purcell}, {Longmore}, {Breen},
  {Green}, {Harvey-Smith}, {Jordan}  \& {Macpherson}}{{Walsh}
  et~al.}{2014}]{2014MNRAS.442.2240W}
{Walsh} A.~J.,  {Purcell} C.~R.,  {Longmore} S.~N.,  {Breen} S.~L.,  {Green}
  J.~A.,  {Harvey-Smith} L.,  {Jordan} C.~H.,   {Macpherson} C.,  2014, \mndoi
  [\mnras] {10.1093/mnras/stu989}, \href
  {http://adsabs.harvard.edu/abs/2014MNRAS.442.2240W} {442, 2240}

\bibitem[\protect\citeauthoryear{{Whiting}}{{Whiting}}{2012}]{2012MNRAS.421.3242W}
{Whiting} M.~T.,  2012, \mndoi [\mnras] {10.1111/j.1365-2966.2012.20548.x},
  \href {http://adsabs.harvard.edu/abs/2012MNRAS.421.3242W} {421, 3242}

\bibitem[\protect\citeauthoryear{{Wright} et~al.,}{{Wright}
  et~al.}{2010}]{2010AJ....140.1868W}
{Wright} E.~L.,  et~al., 2010, \mndoi [\aj] {10.1088/0004-6256/140/6/1868},
  \href {http://adsabs.harvard.edu/abs/2010AJ....140.1868W} {140, 1868}

\bibitem[\protect\citeauthoryear{{Zhang}, {Reid}, {Menten}, {Zheng},
  {Brunthaler}, {Dame}  \& {Xu}}{{Zhang} et~al.}{2013}]{2013ApJ...775...79Z}
{Zhang} B.,  {Reid} M.~J.,  {Menten} K.~M.,  {Zheng} X.~W.,  {Brunthaler} A.,
  {Dame} T.~M.,   {Xu} Y.,  2013, \mndoi [\apj] {10.1088/0004-637X/775/1/79},
  \href {https://ui.adsabs.harvard.edu/abs/2013ApJ...775...79Z} {775, 79}

\makeatother
\end{thebibliography}

\end{document}